\documentclass[11pt] {article} 

\usepackage{times}
\usepackage{graphicx}
\usepackage{supertabular,lscape,epsfig}
\usepackage{amssymb,amsmath}
\usepackage{textcomp}

\numberwithin{equation}{section}

\begin{document}

\title{The local standard of rest and the well in the velocity distribution}

\author{Charles Francis\thanks{Jesus College, Cambridge; e-mail: C.E.H.Francis.75@cantab.net}, Erik Anderson\thanks{360, Iowa St., Ashland, OR 97520, USA.}}

%\email{C.E.H.Francis.75@cantab.net}
\maketitle

\begin{abstract} %ABSTRACT\\
{It is now recognised that the traditional method of calculating the LSR fails. We find an improved estimate of the LSR by making use of the larger and more accurate database provided by XHIP and repeating our preferred analysis from Francis \& Anderson (2009a). We confirm an unexpected high value of $U_0$ by calculating the mean for stars with orbits sufficiently inclined to the Galactic plane that they do not participate in bulk streaming motions. Our best estimate of the solar motion with respect to the LSR $(U_0, V_0, W_0) = (14.1 \pm 1.1, 14.6 \pm 0.4, 6.9 \pm 0.1)$ km\ s$^{-1}$.}
\end{abstract}
{stars: kinematics; stars: statistics; Galaxy: kinematics and dynamics; Galaxy: solar neighbourhood.}

\section{Introduction} \label{sec:Introduction}
\subsection{The meaning of the LSR}
The phrase ``local standard of rest'', or LSR, has two meanings, sometimes distinguished as the dynamical LSR and the kinematical LSR (Delhaye, 1965). The kinematical LSR refers to the mean motion of a specified stellar population. As described in a number of recent papers (e.g., Gontcharov, 2012a \& 2012b; Francis \& Anderson, 2009a ``FA09a''; Dehnen, 1998), kinematics is heavily dependent on stellar age, which in turn leads to dependency on stellar temperature (or colour).

We are here concerned with the determination of the dynamical LSR, which, for the purpose of this paper, will be referred to simply as the LSR. As used here, the LSR is the velocity of a circular orbit at the Solar radius from the Galactic centre, assuming an idealised axisymmetric galaxy in equilibrium, and ignoring features like the bar, spiral arms, and perturbations due to satellite galaxies. It is required to determine dynamical parameters in models of Galactic structure and evolution. Distinct from the kinematical LSR, the dynamical LSR has no dependency on any particular stellar population, but depends only on the gravitational potential field of the Milky Way at the Solar radius.

The central difficulty in calculating the LSR is that while we can observe the kinematics of different stellar populations, these populations are not in circular motion. The traditional method of calculating the LSR, using Str\"omberg's (1924, 1925) asymmetric drift relation, is unreliable because it involves extrapolating a correlation beyond the range of valid data. The rationale for the asymmetric drift relation, as explained in, e.g., Mihalas \& Binney (1981), ch. 6, is that since stars near apocentre come from a denser region of the galaxy, and travel more slowly than stars near pericentre, one expects a greater density of stars near apocentre than near pericentre in any particular region. In consequence, mean transverse orbital velocity should trail the LSR, by an amount known as the asymmetric drift. Greater radial velocities of stars between apocentre and pericentre, are associated with more eccentric Galactic orbits, and greater asymmetric drift.

Following Famaey et al. (2005), \textit{streams} (formerly known as superclusters, Eggen 1958) are here defined as all sky motions. The age and composition of stellar streams was discussed in FA09a and in Francis \& Anderson (2009b ``FA09b'', 2012a ``FA12''). Distinct from streams, Francis \& Anderson (2012b ``XHIP II'') gave a full account of \textit{moving groups} in Hipparcos, i.e. stars sharing a common motion \textit{and} localized in a region of space; \textit{clusters} are gravitationally bound moving groups; \textit{associations} are not gravitationally bound, typically consisting of young stars originating in the same process. The Hyades cluster is thus a different structure from the Hyades stream (though the cluster does belong to the Hyades stream).

The asymmetric drift relation assumes that there are no correlations between stellar orbits, i.e. that the velocity distribution is unstructured or well-mixed (Mihalas \& Binney, 1981); it is sometimes assumed that a well-mixed distribution will result when stellar motions reach steady state (e.g., Binney \& Tremaine, 1987). However, Famaey et al. (2005) observed that the sizes of stellar streams in the solar neighbourhood refute this assumption (it should be noted that well-mixed cannot be identified with dynamic equilibrium, or a centrifuge could not be used to separate substances). It there are streams in dynamic equilibrium, then the distribution is not well-mixed and the kinematic influence of streams on standard statistical measures such as the mean invalidates their use in determining circular motion. 

In practice the majority of stars in the solar neighbourhood belong to one of six clearly defined stellar streams (section \ref{sec:Streams}). FA09b showed that streams are intrinsic parts of Galactic spiral structure and FA12 and Francis (2013) established that around 80\% of disc stars in Hipparcos and in RAVE belong to streams aligned with a bisymmetric spiral, directly refuting the assumptions underlying the use of Str\"omberg's asymmetric drift relation.

The most widely used figure for solar motion with respect to the LSR, \allowbreak$(U_0, V_0, W_0) = (10.00 \pm 0.36, 5.23 \pm 0.62, 7.17 \pm 0.38)$ km\ s$^{-1}$, was calculated by Dehnen \& Binney (1998) using Str\"omberg's relation. However, if $ V<6 $ km\ s$^{-1}$, then over 70\% of the population trail the LSR. In a well-mixed distribution this would mean that stars would spend more than twice as long on the outer parts of their orbits. Velocity at apocentre would be much lower than at pericentre, orbital eccentricities would be much higher than observed, and velocity dispersion would about twice the observed value (appendix \ref{ApB}).

Co\c{s}kuno\u{g}lu et al. (2011) found $V_0 = 13.38 \pm 0.43$ km\ s$^{-1}$ from a separate population of 18\,026 high-probability thin-disc dwarfs within 600 pc of the Sun with data from RAVE, and also using Str\"omberg's relation. Although Co\c{s}kuno\u{g}lu et al.'s result is similar to that of FA09a, the inconsistency with Dehnen and Binney casts further doubt on the validity of the method.

Following the publication of FA09a, Sch\"onrich, Binney \& Dehnen (2010) acknowledged that Str\"omberg's relation breaks down and gave a revised figure $(11.1 \pm 0.7, 12.24 \pm 0.47, 7.25 \pm 0.37)$ km\ s$^{-1}$, with additional systematic uncertainties $\sim(1, 2, 0.5)$ km\ s$^{-1}$. However, they misattributed the non-working of Str\"omberg's relation to the metallicity gradient (section \ref{sec:metallicity}), they calculated $V_0$ using a Galactic model which does not correctly describe the division of the velocity distribution into bulk motions, or stellar streams, and they assumed that $U_0 = - \overline{U}$, whereas, as observed by Famaey et al. (2005), the size of the stellar streams in the solar neighbourhood invalidates the calculation of the LSR from mean motions (the mean is stated in the paper, but in a review Sch\"onrich has said that he used an undocumented average. The argument remains that any average is invalidated if it does not take account of streaming motions). Subsequently Sch\"onrich (2012) revised $U_0$ upwards to match the value found here, but without giving clear justification.

\subsection{The well in the velocity distribution}
Since it is not possible to estimate the LSR from mean motions, it is necessary to study the features of the local velocity distribution. FA09a observed that the velocity distribution contains a deep trough. They suggested that a minimum might be expected at the LSR as a consequence of disc heating. Heating is the hypothetical process by which scattering events cause the random velocities of stars to increase with age (e.g., Jenkins, 1992). 

The Maxwell distribution for molecular speed assumes that thermal motions in each velocity component are Gaussian and independent. This is reasonable in a gas because molecular collisions are elastic and the typical time scale between them is of the order of nanoseconds, so that there is a rapid and essentially random transfer of momentum between molecules. The dynamics of stellar orbits are quite different. For orbital motions of stars the radial and azimuthal components are not independent; the magnitude of the radial component is greatest when magnitude of the azimuthal component is small, and the magnitude of the azimuthal component has maxima at the apsides, when the radial component is zero. Momentum transfer between orbits is slow. In a steady axisymmetric potential, circular motion is a minimum of energy at fixed angular momentum. Thus zero temperature, or minimum kinetic energy, describes a system in which all bodies follow circular orbits. A heating mechanism, including mixing by dynamical friction, in which momentum is transferred from `hot', high eccentricity, orbits to those of low eccentricity, would result in a systematic movement away from circular motion. As a result the distribution in velocity space might be expected to have a minimum at circular motion.

Subsequently, FA09b showed that the true cause of the minimum is the alignment of orbits to the spiral potential of the arms, as was previously suggested by the stellar migration hypothesis in the form given by L\'{e}pine et al. (2003) and discussed in section \ref{sec:Streams}. In practice FA09a found a deep trough in the vicinity of $V=-12$ km\ s$^{-1}$, containing a number of minima and it was not possible to be certain that the chosen candidate gave the correct value of $U_0$. FA09a gave as an estimate for solar motion $ (7.5 \pm 1.0, 13.5 \pm 0.3, 6.8 \pm 0.1) $ km\ s$^{-1}$, based on the choice of a particular minimum, which appeared to be the most marked, and which was in best agreement with previous estimates. There is some difficulty in the method, because the presence of moving groups with velocities near to that of the LSR can obscure the correct minimum.

\subsection{XHIP}
The Extended Hipparcos Compilation (``XHIP'', Anderson \& Francis, 2012) gives radial velocities for $ 46\,392 $ Hipparcos stars, together with recalculated memberships of moving groups and multiplicity information from the Catalog of Components of Double \& Multiple Stars (Dommanget \& Nys, 2002) and The Washington Visual Double Star Catalog, (Mason et al., 2001, version 2010-11-21). XHIP has been tested for kinematic bias. It is found that any bias is minimal compared to the size of the population. The method used here to calculate the LSR is in any case not influenced by kinematic selection biases. Stars in clusters and associations may be over-represented. Even without selection bias, these stars produce unwanted peaks in the distribution which may distort analysis. They were removed from the population. We limited the database to $ 26\,103 $ stars, by removing stars outside 300 pc, stars with parallax errors greater than 20\%, stars with quoted radial velocity errors greater than $5$ km\ s$^{-1}$, stars with quality index ``D'' (which includes unsolved binaries), secondary stars in multiple systems, stars in moving groups, and stars with velocities greater than 350 km s$^{-1} $ relative to the Galactic centre, taken as $(-14, -225, -7)$ km\ s$^{-1}$ (the last cut removes seven stars).

Fifteen hundred stars used by FA09a were excluded, either because they belong to a moving group, or because of improved identification of multiple systems, so that the new database contains seven thousand new members with complete kinematic information within 300 pc. A larger number have revised radial velocity data, because of more recent measurement, because of the use of weighted mean velocities when there are a number of measurements, and because of systematic calibration corrections made in the Pulkovo compilation of radial velocities (Gontcharov, 2006).

Given this larger database, with more accurate radial velocities, memberships of moving groups and improved multiplicity information, it is appropriate to recalculate the LSR. It transpires that the wrong minimum was chosen in FA09a for circular motion, and that a substantially greater value of $ U_{0} $ should have been given. We have confirmed this value by an independent calculation of mean velocity for stars with orbits substantially inclined to the Galactic plane.

\subsection{Galactic Parameters}
Francis \& Anderson (2013) compiled a database of over 150 determinations of the distance to the Galactic centre, and found that, despite a number of recent high profile estimates against the trend, the historical downward trend in estimates of $ R_0 $ has continued this century. The mean of 48 estimates since 2000 is 8.0 kpc, but for 19 estimates published since 2010 the mean is $ R_0 = 7.9 $ kpc. The downward trend may indicate that estimates remain a little high. They also found that a number of estimates from luminosity distance are high, because the mean magnitude of the red clump in Hipparcos stars is fainter than the peak magnitude, and because the use of non-standard reddening in some determinations is undermined by the calculation of Girardi and Salaris (2001), who show that red clump stars in the bulge are intrinsically redder than local counterparts. They gave new calculation from the halo centroid using revised cluster distances ($ 7.4 \pm 0.2|_{\mathrm{stat}} \pm 0.2|_{\mathrm{sys}} $), and a calculation from red clump stars in the bulge with 2MASS magnitudes ($ 7.5 \pm 0.3 $ kpc), calibrated to Hipparcos parallax distances and taking population effects into account. Using this calibration agreement was found with eight other determinations of $ R_0 $ from the red clump. For this paper we adopted $ R_0 = 7.5 $ kpc. We used solar velocity 227 km\ s$^{-1}$ in the direction of Galactic rotation, found from measurement of the proper motion of Sgr A* (Reid and Brunthaler, 2004) under the assumption that Sgr A* is stationary at the Galactic centre.

The values of $ R_0 $ and $ V_{\odot} $ are not critical to the calculations in this paper; the use of higher adopted values would make almost no impact on our result. Our adopted value for $ R_0 $ agrees with Sofue et al (2011) and Bobylev (2013) who used data from a number of populations of circular motion tracers and found respectively $ R_0 = 7.54 \pm 0.77 $ kpc and a combined estimate $ R_0 = 7.5 \pm 0.3 $ kpc. Our adopted value also agrees with $ 7.72 \pm 0.33 $ kpc given by Gillessen (2009) from the orbit of S2 after excluding data from 2002 when S2 was near pericentre. We believe this data (which leads to a higher estimate of $ R_0 $) should be excluded because the method is highly sensitive to unmodelled effects such as distributed mass in the bar and the relativistic Lense-Thirring effect, or frame dragging, due to the rotation of Sgr A*.

For an elliptical orbit the eccentricity vector is defined as the vector pointing toward pericentre and with magnitude equal to the orbit's scalar eccentricity. It is given by
\[ \mathbf{e}=\frac{|\mathbf{v}|^{2}\mathbf{r}}{\mu} - \frac{(\mathbf{r}\cdot\mathbf{v})\mathbf{v}}{\mu} - 
\frac{\mathbf{r}}{|\mathbf{r}|}.
\]
where $\mathbf{v}$ is the velocity vector, $\mathbf{r}$ is the radial vector, and $\mu=GM$ is the standard gravitational parameter for an orbit about a mass $M$, (e.g. Arnold, 1989; Goldstein, 1980). We adopted $ \mu=R_0 \Theta_0^2 = 1.04 \times 10^{22} $ km$^3 $ s$ ^{-2} $, where $ \Theta_0=212 $ km s$ ^{-1} $ is the speed of circular motion, based on our adopted value for $ V_{\odot} $ and our calculated value of $ V_0 $.

For a Keplerian orbit the eccentricity vector is a constant of the motion. Stellar orbits are not strictly elliptical, but rosette orbits can usefully be regarded as precessing ellipses and the eccentricity vector remains a useful measure (the Laplace-Runge-Lenz vector, which is the same up to a multiplicative factor, is also used to describe perturbations to elliptical orbits). The eccentricity vector describes the osculating Kepler orbit, i.e. the orbit a star would have if distributed Galactic mass were replaced by a constant equivalent central mass. The osculating orbit thus matches the actual orbit in both first and second derivatives (velocity and acceleration). The above value for $ \mu $ is equivalent to an effective central mass of 79 billion $ M_{\odot} $. Generally, for an axisymmetric mass distribution the effective central mass varies with Galactic radius and the eccentricity vector will precess, but FA09b showed that perturbations due to spiral arms tie the rate of precession to spiral pattern speed. FA09a and Francis (2013) found from the alignment of the orbits of Hipparcos and RAVE stars with more distant spiral structure seen in 2MASS and in gas distributions that spiral pattern speed is small.

\section{Determination of the local standard of rest}\label{sec:LSR}
\subsection{Determination of $ W_{0} $}
\begin{figure}
	\centering
	\includegraphics[width=1\textwidth]{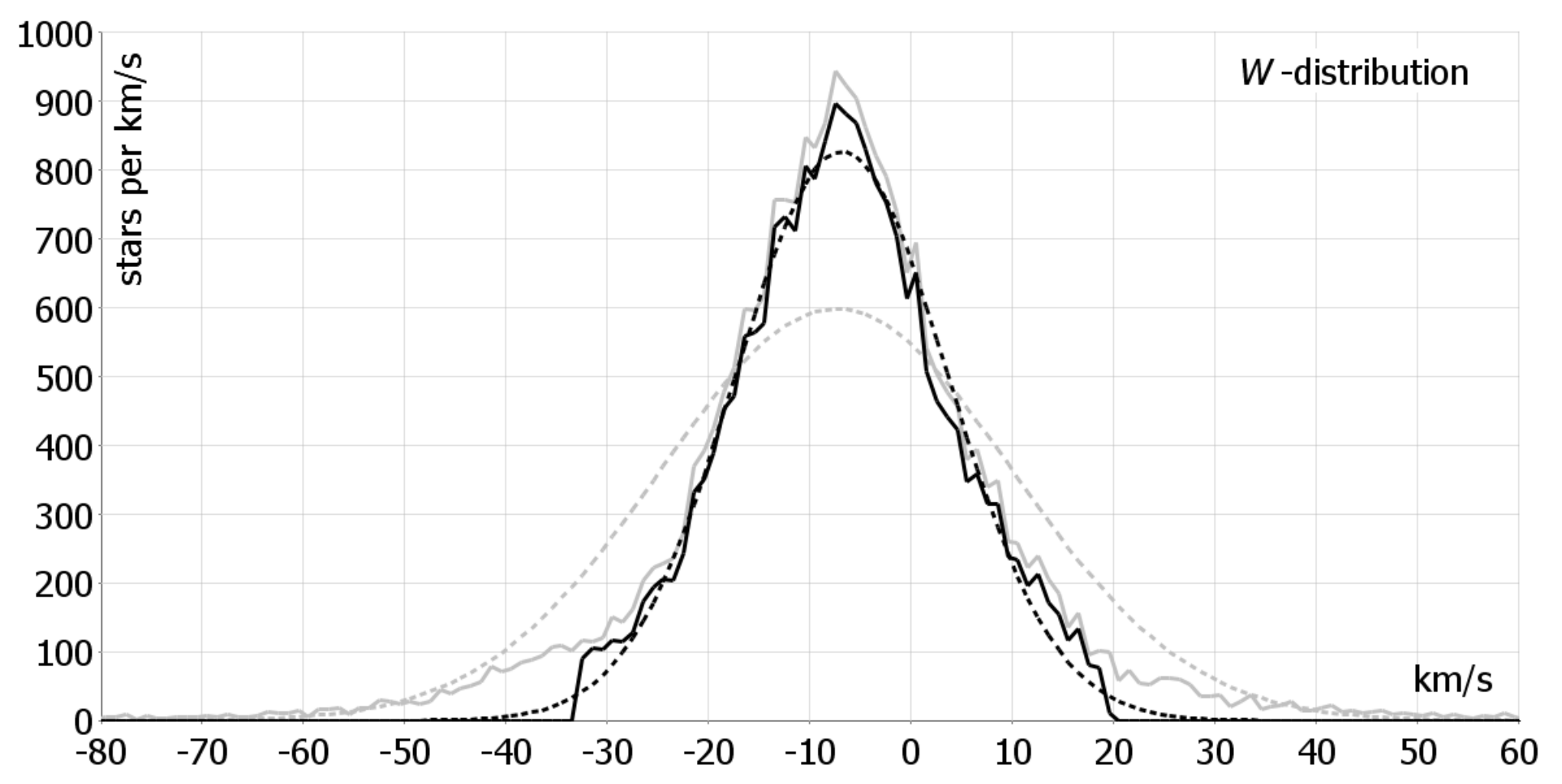}
	\caption{The velocity distribution perpendicular to the Galactic plane for the entire population (continuous, grey) and after truncating to find a best fit with a Gaussian distribution (continuous, black). Gaussian distributions with the same mean and standard deviation are shown as dotted lines.}
	\label{fig:wdist2}
\end{figure}

Ignoring the possibility of perturbations to the Galactic plane, motions of thin disc stars in the $W$-direction may be treated as a low amplitude oscillation due to the gravity of the disc, and as independent of orbital motion in the $U$-$V$ plane. Calculating the mean $W$-velocity is problematic, because outliers have a disproportionate effect on the mean. Simply discarding velocities a given number of standard distributions from the mean is unreliable, because values opposing any error in the mean will be preferentially removed, resulting in a compounded error and potentially leading to non-convergence on iteration of the method. To find $W_0$, we first removed stars with eccentricity in the $ UV $ plane greater than $0.4$, based on $(U_0, V_0) = (14.1, 14.6)$ km s$^{-1}$. We then restricted the population to stars with $|W + W_{\mathrm{G}}| < w$, and varied $W_{\mathrm{G}}$ and $w$ so as to minimise the sum of least squares differences between the frequency distribution, in $1$ km\ s$^{-1}$ bins, and a Gaussian distribution with mean $W_{\mathrm{G}}$ and standard deviation equal to that of the population. The best fit (figure \ref{fig:wdist2}) has $W_{\mathrm{G}} = 6.9 \pm 0.1$, $w = 26 \pm 1$ and gives an estimate for the component of the LSR perpendicular to the disc, $ W_0 = - \overline{W} = 6.9 \pm 0.1 $. The cut-off is at $ 2.5 $ standard deviations of the Gaussian distribution ($ 10.5 $ km s$^{-1}$). 

As described by Robin et al. (2003), thick disc stars constitute 6.8\% (by number) of the local stellar density, corresponding to a thin-disc boundary at \linebreak$ | W + 6.9 | \approx 31 $ km s$ ^{-1} $, in agreement with 3 standard deviations of the best fit Gaussian distribution. Bovy et al. (2012) observed no evidence of a crisp division into distinct populations in SEGUE, and Francis (2013) found a unimodal distribution of metallicities for thin and thick disc stars in RAVE. The thin-disc boundary is thus a matter of convenience and not a physical division. Robin et al.'s figure for proportion of stars in the thick disk is high because they cite studies (like our own) of Hipparcos stars. It includes a significant number of halo stars, which are overrepresented in Hipparcos because they were chosen for the input catalogue. The number of thick disc stars may be inflated for the same reason, but this has negligible impact on the best fit Gaussian.

In a first approximation using planar motion, a star in the Galactic plane in a circular orbit with velocity $ 31 $ km s$ ^{-1} $ perpendicular to the plane will rise to a height of $ 1.1 $ kpc above the plane (using $ R_{0} = 7.4 $ and velocity of circular motion $ 212 $ km s$ ^{-1} $) a factor of $ \sim 4$-$8 $ greater than typical scale heights given for the thin disc in the literature (e.g., Gilmore and Reid, 1983, Robin et al., 2003). This is an overestimate, but since the total gravitating mass of the inner Galaxy and halo greatly outweighs matter distributed in the disc at the solar radius, it may not be a substantial overestimate.

\subsection{The well in the $ U$-$V $ distribution}
It is sometimes thought to fit Gaussians to the $ U $ and $ V $ distributions, as was done in FA09a. There is no reason to repeat this analysis; stellar motions are two dimensional in the Galactic plane, and orbit the LSR in the $ UV $-(velocity) plane), meaning that there is a strong relationship between $ U $ and $ V $ such that  one dimensional Gaussian fits for $ U $ and $ V $ are misleading.

\begin{figure}
	\centering
			\includegraphics[width=1\textwidth]{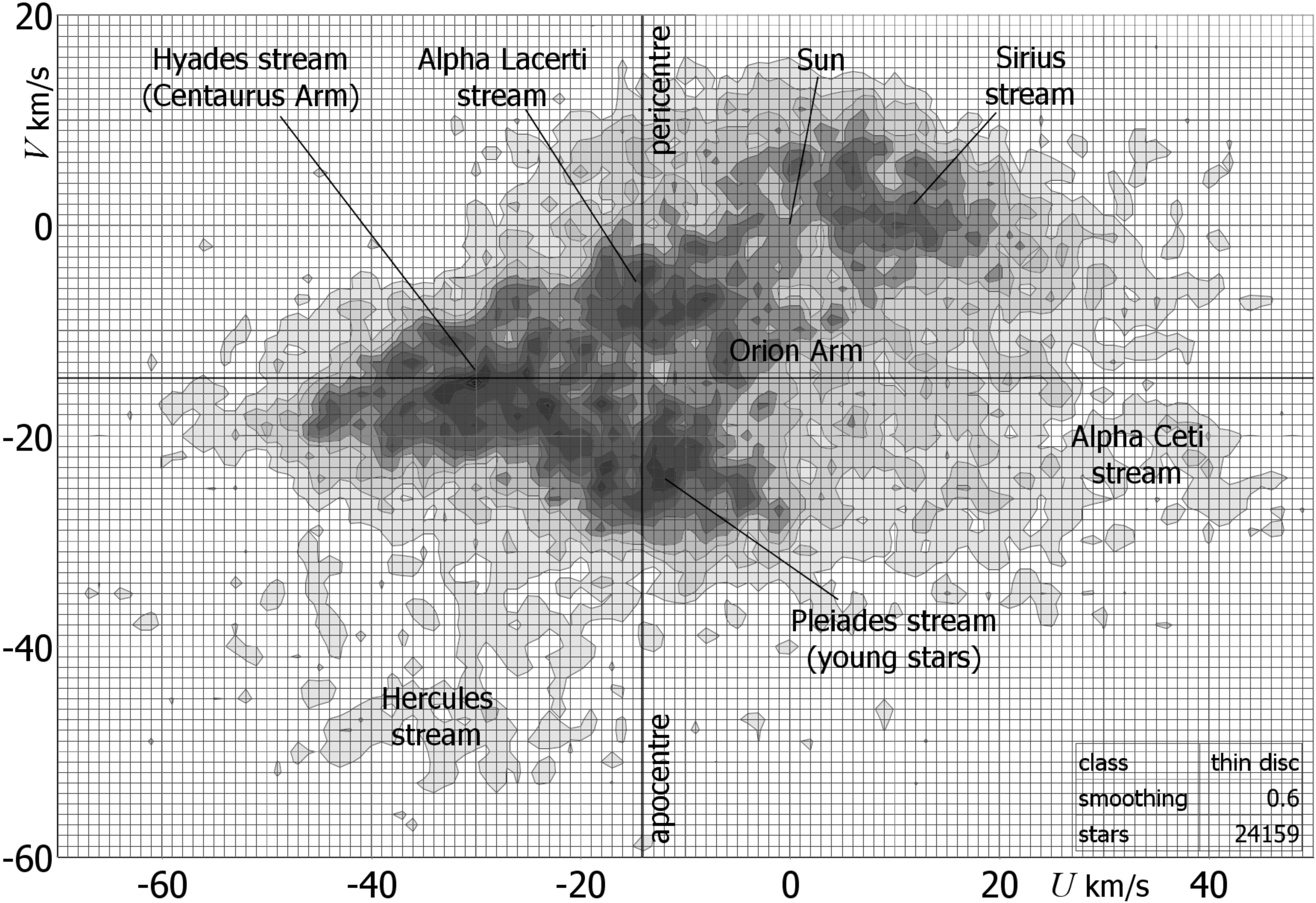}
	\caption{The distribution of $U$- and $V$-velocities using Gaussian smoothing with a standard deviation of $0.6$ km\ s$^{-1}$, showing the Hyades, Pleiades, Sirius, Hercules, Alpha Lacertae and Alpha Ceti streams (from ). Axes are shown with an origin at the LSR, $(U_0, V_0) = (14, 14.5)$ km\ s$^{-1}$. Moving groups (localised in space as well as velocity), non-primary stellar components, and stars with $| W + 6.9| > 31$ km\ s$^{-1}$ are excluded.}
	\label{fig:UVdist}
\end{figure}

We restricted the velocity distribution to 24\,246 thin disc stars with \linebreak$ | W + 6.9 | \leq 31 $ km s-1 and plotted the velocity distribution for the population (figure \ref{fig:UVdist}), using Gaussian smoothing (see, e.g., FA09a) with standard deviation 0.6 km\ s$^{-1}$. The larger population, and the removal of clusters and associations (especially the Hyades cluster) using group memberships determined in XHIP II, gives a clearer image of the features of the distribution than was previously available. The Hyades, Pleiades, Sirius, Hercules, Alpha Lacertae and Alpha Ceti streams are clearly visible, with some overlap between streams. Figure \ref{fig:UVdist} shows a clearly defined, and unique, central well at $(U, V) = (-12.5, -14)$ km\ s$^{-1}$. The statistical significance of the well is at least 99.8\% (appendix \ref{ApA}). This is expected to be close to the LSR, but is not expected to be the best estimate of the LSR because the population contains stars at a range of radii from the Galactic centre. The method described below removes the effect of differential rotation by calculating orbital eccentricity in order to determine circular motion.

The distributions of stars within 50 pc of the Galactic plane and at greater than 50 pc from the Galactic plane are similar to figure 2, but showing greater random fluctuations as is to be expected. Francis (2013) found that major features, including the central well, are visible in the population of RAVE stars and much greater distances from the Galactic plane. We may conclude that the well is not an artefact, caused, for example, by dust in the Galactic plane. 

\subsection{Determination of $ (U_{0}, V_{0}) $}

Famaey et al (2005) studied the kinematics of a population of mainly giants determined by magnitude and colour. Inevitably this population contains a subpopulation of pre-main sequence stars on Hayashi tracks, which they identified as having young star kinematics. Francis (2013) showed peaks in the distribution at close to the LSR for populations containing substantial numbers of pre-main sequence stars in Hipparcos and RAVE. Young stars have velocities dependent on the kinematics of the gas clouds from which they are formed, and are likely to have motions closer to the LSR than the norm. They will also have a random peculiar motions (e.g., Gontcharov, 2012a), which will appear as noise in our attempt to determine the position of the well in the velocity distribution of mature stars. Stars in associations have already been removed, and we also removed stars with $B-V \le 0.3$ mag. The removal of stars with $B-V \le 0.3$ mag clarifies the position of the well in the velocity distribution. It does not create it; these stars have not been removed from figure \ref{fig:UVdist}, which shows the well. The well in the distribution is also seen at close to our current estimate of the LSR in figure 8 of FA09a, with greater or lesser clarity in the velocity distributions for each colour band, including those for hot stars. It is unfortunate that at the time of writing of that paper, we did not recognise the significance of this feature of those plots, or we might not now have to correct our estimate.

\begin{figure}
	\centering
	\includegraphics[width=1\textwidth]{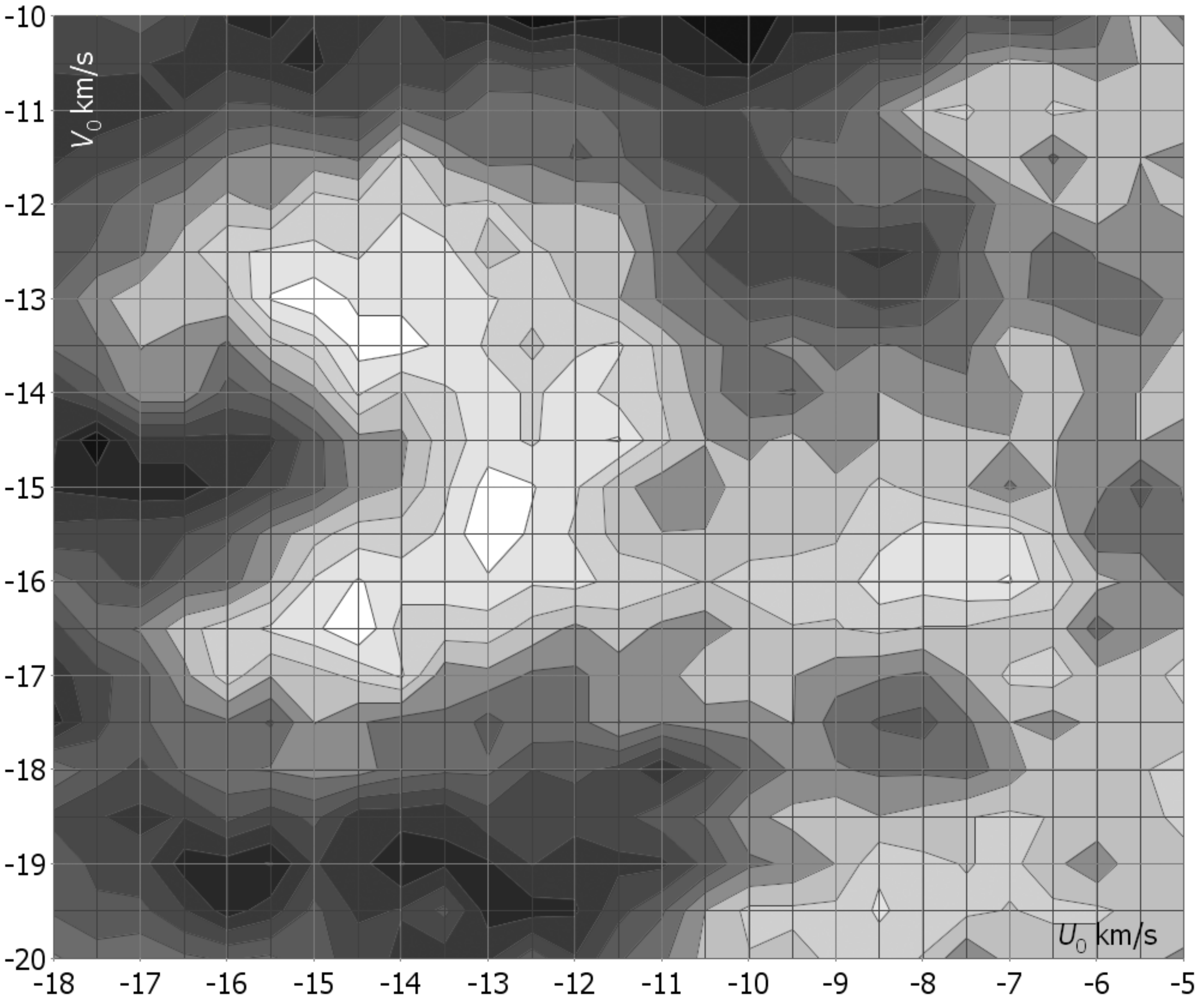}
	\caption{Contour of the numbers of stars with eccentricity less than 0.01 for different values of $(U_0, V_0)$, in multiples of 0.5 km\ s$^{-1}$. Stars with $B-V \le 0.3$ mag, stars in moving groups, non-primary stellar components, and stars with $| W + 6.9| > 31$ km\ s$^{-1}$ are excluded. Values range from 18 to 54 stars.}
	\label{fig:circularorbits}
\end{figure}

We plotted the number of stars with eccentricity less than 0.01 for different values of $(U_0, V_0)$, in multiples of $1$ km\ s$^{-1}$ (figure \ref{fig:circularorbits}). We found a deep well containing three minima in the vicinity of $(-14, -14.5)$ km\ s$^{-1}$. It appears that the true position of the LSR is obscured by the peak at $(-17, -15)$ km\ s$^{-1}$. This peak may be interpreted as being due to stars from a particular star forming process, leading to a moving group which is now too dispersed in space to be identified as an association, but which is too recent for orbital alignment with the arms. Ignoring this peak the expected position of the minimum can be gauged from the position of the sides of the well, giving an estimate for the LSR, $(U_0, V_0) = (14.0 \pm 1.2, 14.5 \pm 1.0)$ km\ s$^{-1}$.
\begin{figure}
	\centering
				\includegraphics[width=1\textwidth]{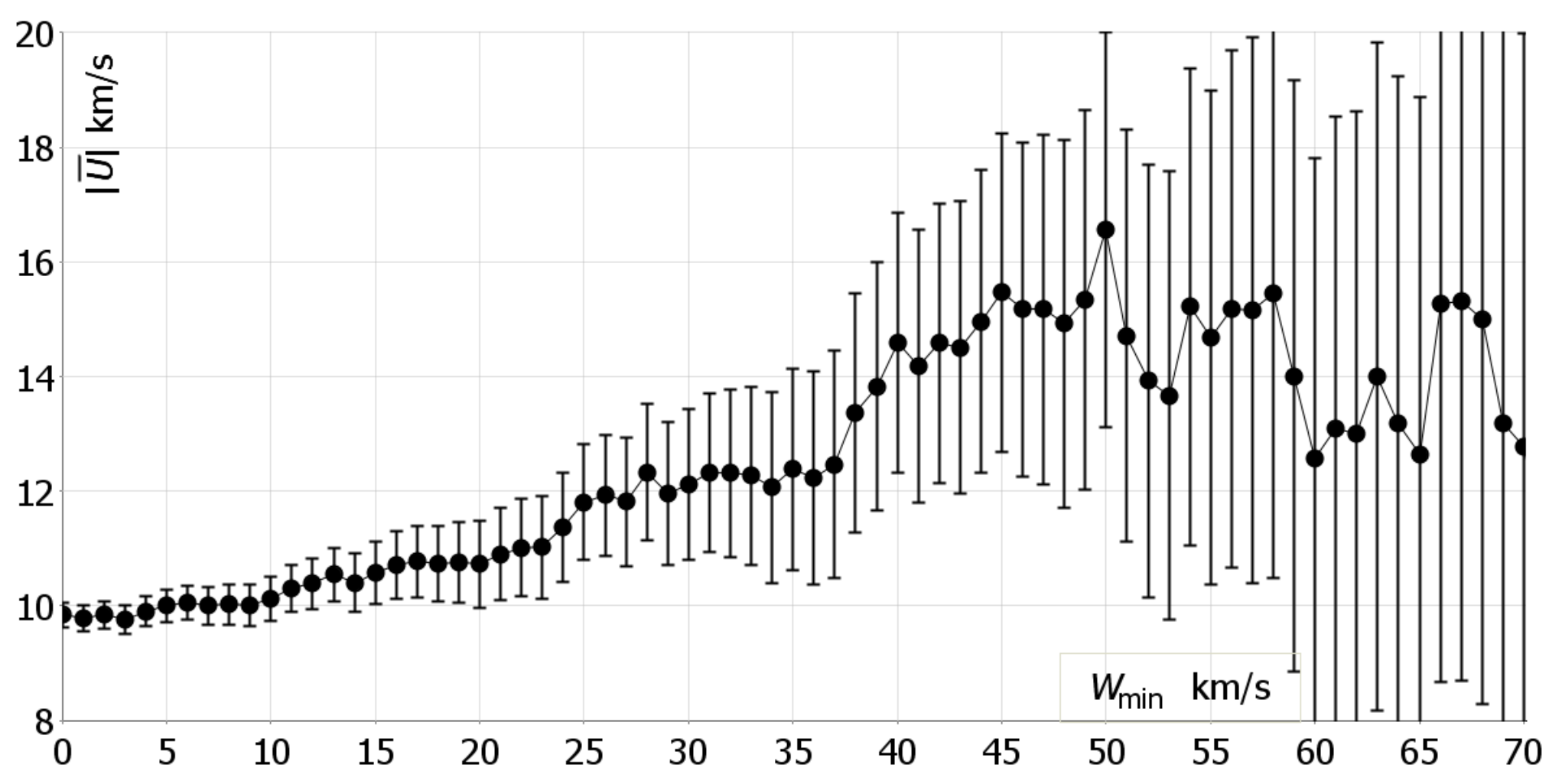}
	\caption{Mean $ U $ velocities, for stars with $ |W+6.9|>W_{\mathrm{min}} $.}
	\label{fig:Uthick}
\end{figure}

This value for $U_0$ is substantially greater than previous estimates. Although bulk motions in the Galactic plane invalidate any calculation of $U_0$ and $V_0$ from the mean of a population of disc stars, it remains legitimate to calculate $U_0$ from the mean of stars with orbits sufficiently inclined to the Galactic plane that they do not participate in spiral structure and are not members of a stellar stream (Klement et al. 2011 found no evidence of significant streams in the halo), which was found dependent on the misidentification giants as dwarfs. It is found that $ |\overline{U}| $ increases with cut on the component of velocity perpendicular to the Galactic plane, up to $ W_{\mathrm{min}}=39 $ km\ s$^{-1}$, above which it shows random fluctuation (figure \ref{fig:Uthick}), indicating that the influence of bulk streaming motions is no longer significant. For a population of 841 stars within 300 pc, with parallax errors less than 20\%, and with velocity, $W$, perpendicular to the Galactic plane with $| W + 6.9 | \ge 40$ km\ s$^{-1}$, we found mean velocity $\overline{U} = -14.6 \pm 2.3$ km\ s$^{-1}$, in excellent agreement with the estimate from stars in circular motion. Since the methods are distinct and based on different populations, they can be combined to give a best estimate $U_0 = 14.1 \pm 1.1$ km\ s$^{-1}$.

\begin{figure}
	\centering
				\includegraphics[width=1\textwidth]{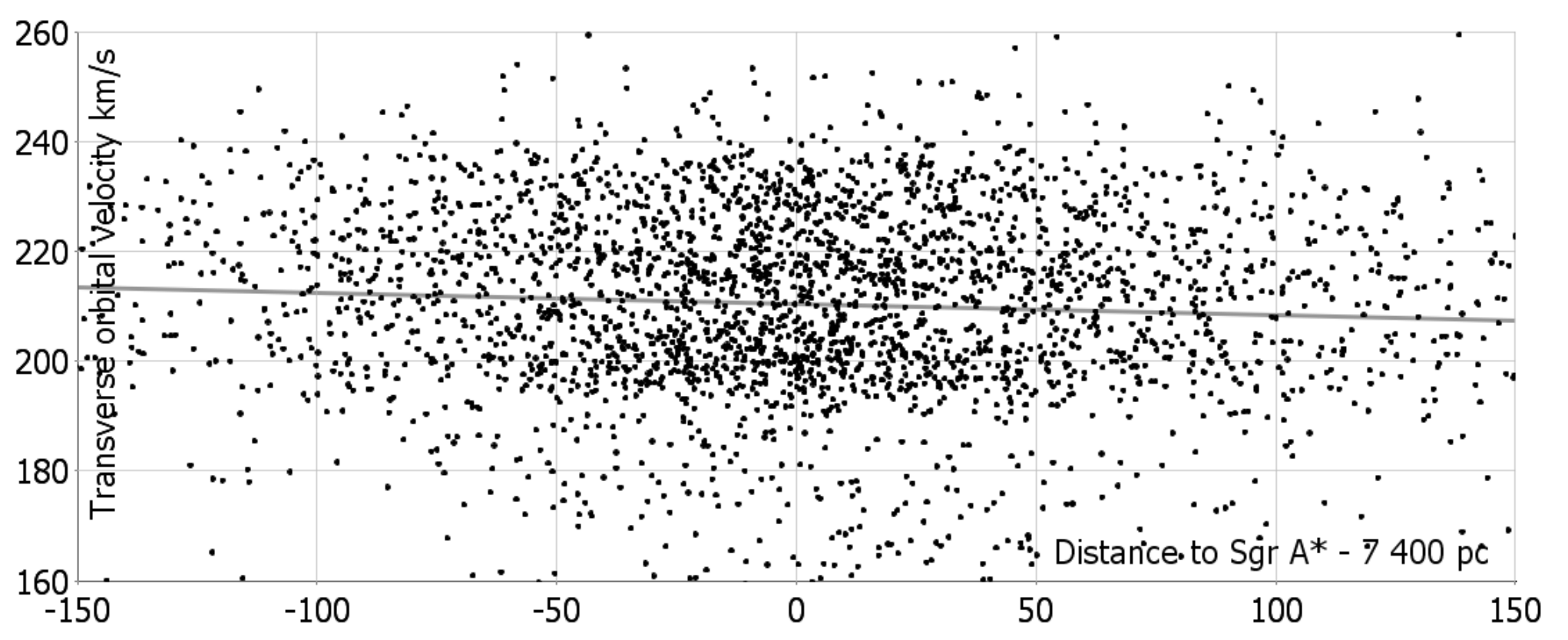}
	\caption{Transverse orbital velocity plotted against distance to the Galactic centre for stars with $| U + 10 | < 6$ km\ s$^{-1}$ and $ B-V > 0.3$ mag. The circular speed curve is seen as a trough in the distribution of with a gradient of $\sim -15$ km\ s$^{-1}$ kpc$^{-1}$.}
	\label{fig:6}
\end{figure}

\begin{figure}
	\centering
				\includegraphics[width=1\textwidth]{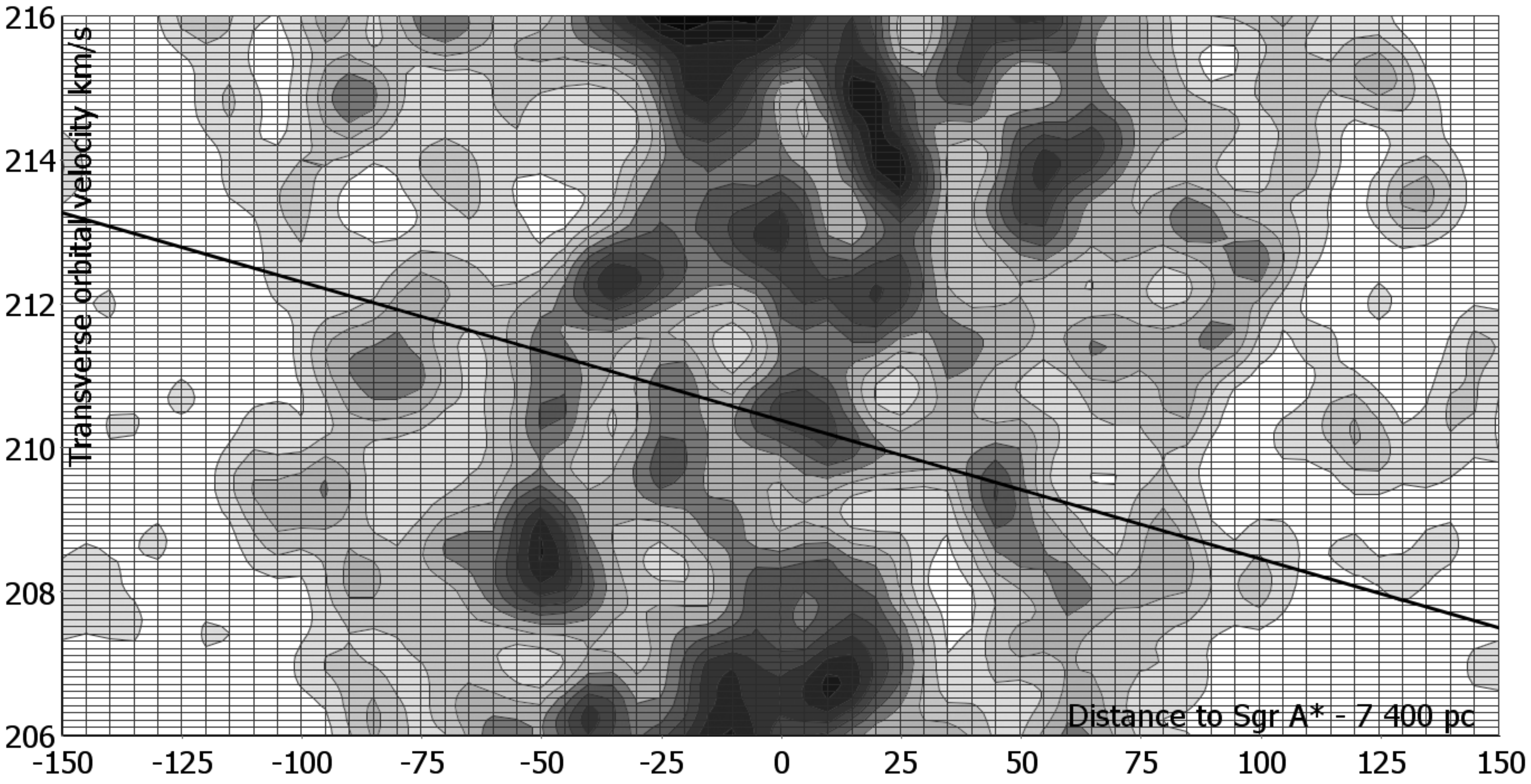}
	\caption{The transverse velocity distribution of stars with $| U + 10 | < 6$ km\ s$^{-1}$ and $206<V_{\mathsf{T}} <216 $ km\ s$^{-1}$, excluding stars with $ B-V \le 0.3$ mag, using Gaussian smoothing with parameters $\sigma_{v} = 0.4$, $\sigma_R = 4$. The line of regression through the minima at constant distance is also shown. At the solar radius, the minimum is at $V_{\mathsf{T}} = 210.4 \pm 0.4$ km\ s$^{-1}$.}
	\label{fig:7}
\end{figure}

We confirmed the value of $V_0$ by plotting transverse orbital velocity, $V_{\mathsf{T}}$, against distance to the Galactic centre for stars with $ B-V > 0.3$ mag and \linebreak$| U + 10 | < 6$ (figure \ref{fig:6}). This interval gives a sufficient population for a significant result, but avoids the moving group at $ (-17, -15) $ km\ s$^{-1}$. There is a clear trough in the distribution with gradient $\sim -15$ km\ s$^{-1}$ pc$^{-1}$, demarcating the circular speed curve. To estimate the position of the minimum we smoothed the distribution using Gaussian smoothing (figure \ref{fig:7}). Using different values of the smoothing parameters $\sigma_{v} = 0.3, 0.4, 0.5, 0.6, 0.7$, $\sigma_R = 2, 3, 4, 5, 6$, we determined the regression line through the points of minimum density at given distance, finding a slope $-15.6 \pm 4.3 $ km\ s$^{-1}$ kpc$^{-1}$, consistent with the gradient at the solar radius shown from gas distributions by Clemens (1985), who also shows that the rotation curve is flat on the large scale. The intercept at the solar radius is $210.4\pm 0.4$ km\ s$^{-1}$, in agreement with the value found by restricting eccentricity to $ 0.01 $. Because of similarities in the method, it is not permissible to use a weighted mean estimate. It is legitimate to use the narrower error margin.


\section{The reason for the central well}\label{sec:Streams}
\begin{figure}
	\centering
		\includegraphics[width=1\textwidth]{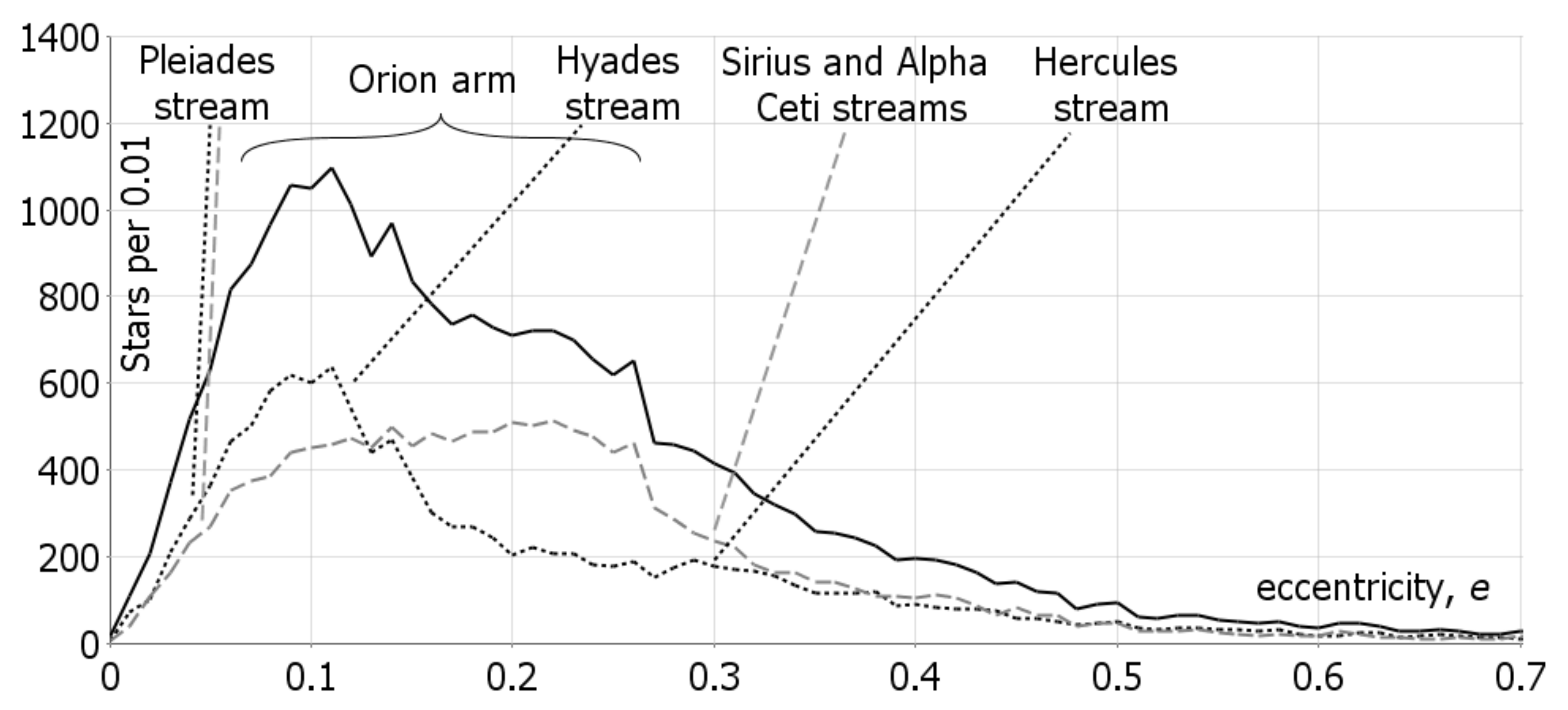}
	\caption{The eccentricity distribution. The majority of stars have eccentricities in the range $0.07<e< 0.27$, consistent with the variation in orbital radius predicted by L\'{e}pine. Stars on inbound orbits are shown by dashed lines, and on outbound orbits by dotted lines. The whole population is shown by a continuous line. Calculation of eccentricity is dependent on the LSR, but the eccentricity distribution is qualitatively unchanged for a wide range of values for the LSR.}
	\label{fig:eccdist}
\end{figure}

According to the stellar migration hypothesis in the form given by L\'{e}pine et al., stars do not follow circular orbits, but are perturbed by the gravity of a spiral arm such that orbital radius varies by about 2-3 kpc over periods in the order of 1 billion years. This prediction is confirmed by the deep well seen in the velocity distribution at the position of circular motion (figure \ref{fig:UVdist}), and also by the eccentricity distribution (figure \ref{fig:eccdist}), which shows the bulk of orbits with eccentricities between 0.07 and 0.27. An earlier form of stellar migration (Sellwood and Binney, 2002), in which stars on near circular orbits migrate slowly because of the action of spiral waves, is not supported by the observed velocity distribution, as orbits are not near circular.
\begin{figure}
	\centering
		\includegraphics[width=1\textwidth]{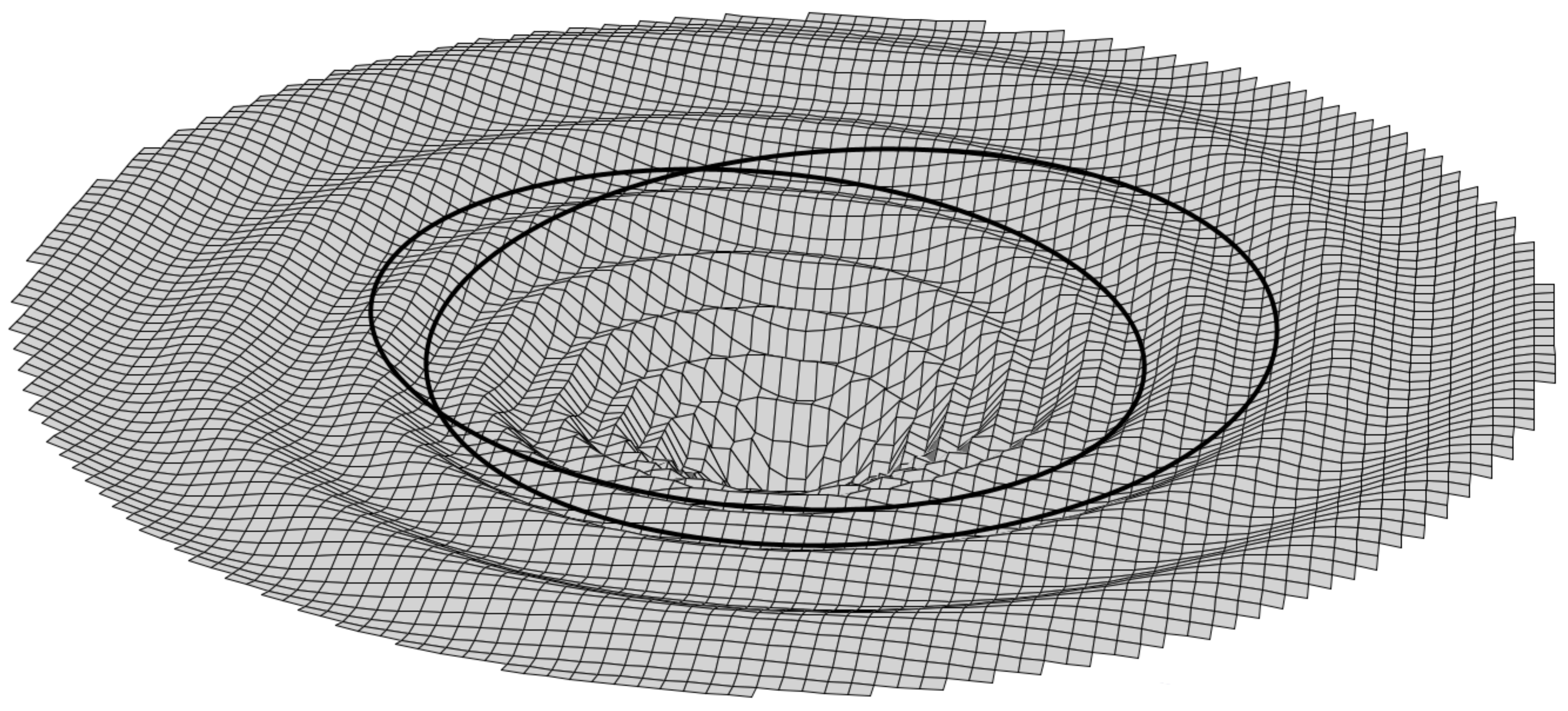}	
		\caption{The gravitational potential of a bisymmetric spiral galaxy plotted on a vertical axis against the galactic disc on a horizontal plane. The alignment of (idealised) elliptical orbits with troughs in the potential is shown.}
	\label{Fig:18}
\end{figure}

A mechanism for the alignment of orbits with the arms can be understood by plotting the gravitational potential for a spiral galaxy (vertical axis) against the plane of the disc (figure \ref{Fig:18}). Stellar orbits in such a potential are precisely analogous to the trajectories of particles in a frictionless spiral grooved funnel in a uniform gravitational field, for which potential is directly proportional to height. A particle at the highest point of its path, where it is moving least quickly, will tend to fall into a groove and then follow the groove downwards, picking up speed as it goes. Eventually, the particle gains enough momentum to jump free of its groove. It crosses over the next-highest groove (for a bisymmetric spiral), then falls back to a higher point in its original groove. Thus orbits follow the arms on the inward part of the motion. The tendency for stars to follow the arms reinforces the gravitational potential of the arm. Such a model is insensitive to the shape of the funnel, can account for the observed frequency of spirals in galaxies with a wide range of sizes and mass distributions, and explains star formation in spiral arms as a consequence of collisions between gas clouds travelling inward along one arm with clouds travelling outwards, having left the other arm (FA09b). 

\begin{figure}
	\centering
		\includegraphics[width=1\textwidth]{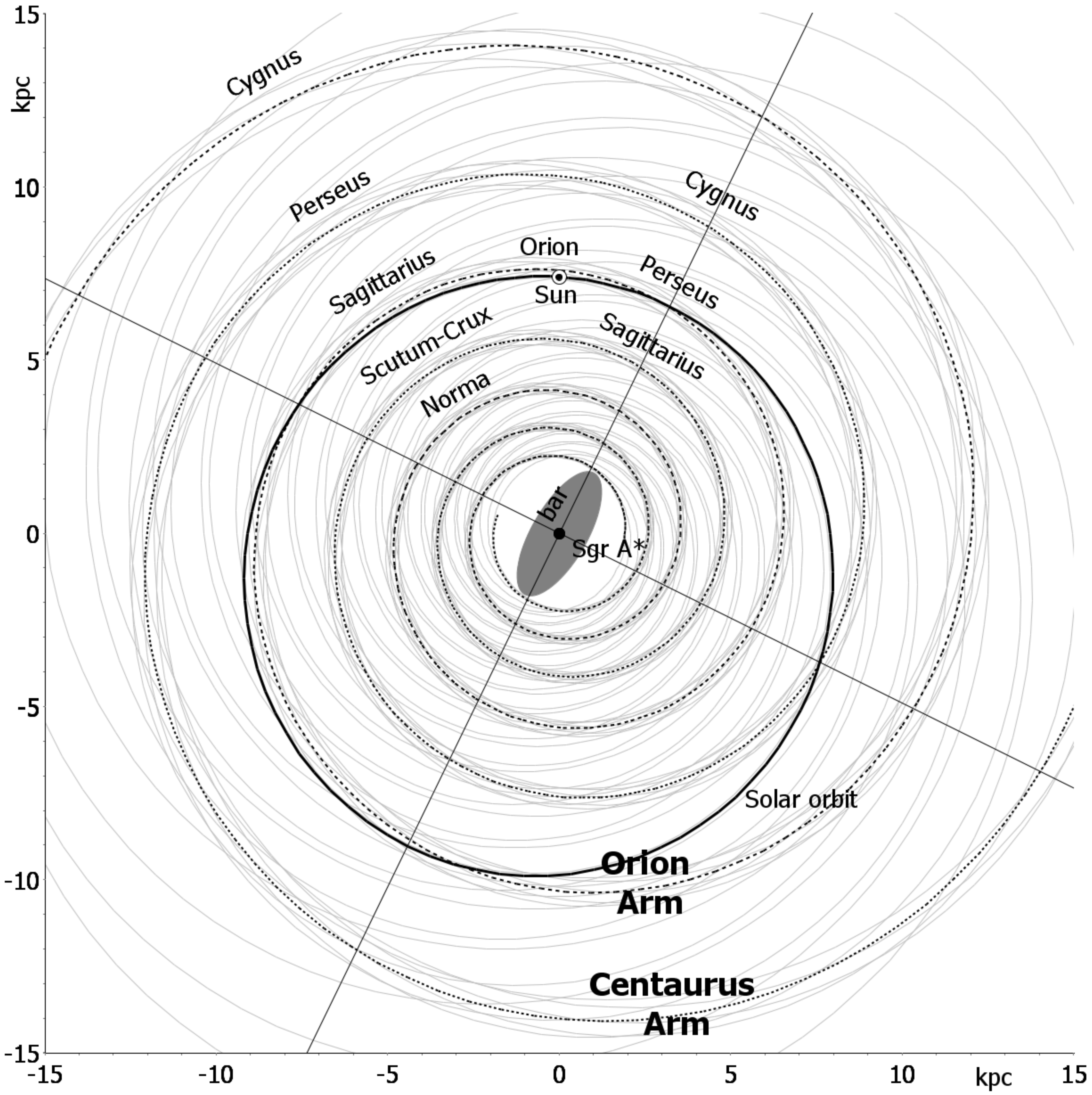}
	\caption{Axisymmetric logarithmic double spiral with pitch angle 5.56° constructed from ellipses of eccentricity 1.6 (from FA12). The plot uses solar velocity $ 225 $ km s$ ^{-1} $ and distance to the Galactic centre 7.4 kpc, but figures are scalable. The solar orbit is shown in approximation (neglecting precession), together with its major axis and latus rectum, based on solar motion $ (U_{0}, V_{0}) = (14, 14.5) $  km s$ ^{-1} $.}
	\label{fig:9}
\end{figure}

The alignment of orbits to spiral structure can seen by repeatedly enlarging an ellipse by a constant factor, $ k $, centred at the focus and rotating it by a constant angle, $ \tau $, with each enlargement (figure \ref{fig:9}). The pitch angle of the spiral depends on $ k $ and $ \tau $, not on the eccentricity of the ellipse, but, for a given pitch angle, ellipses with a range of eccentricities can be fitted to the spiral, depending on how narrow one wants to make the spiral and what proportion of the circumference of each ellipse one wants to lie within it. Higher eccentricity orbits fit spirals with higher pitch angles. Thus stars move along an arm on the inward part of their orbits, leaving the arm soon after pericentre, crossing the other arm on the outward part and rejoining the original arm before apocentre. In this description of material arms there is no winding problem because the spiral depends on the \textit{paths} of stellar orbits, not on orbital velocity.

FA12 found that a bisymmetric spiral fits distributions of neutral and ionised gas, and also fits the observed distribution of peaks in the Galactic plane in data from 2MASS. They found that nearly eighty percent of local disc stars have orbits aligned with the spiral. With the fit to a two armed ``grand design'' spiral, it is found that the so-called Orion spur is not a spur, but is a part of one of two major spiral arms. We have called this the Orion arm. We have called the other arm the Centaurus arm. The section of the Orion arm in the direction of rotation was formerly thought to be part of the Perseus arm, while that in the opposite direction was thought to be part of the Sagittarius arm. Similarly the Centaurus arm inward of the Sun contains sections which were thought to be parts of the Sagittarius and Scutum-Crux arms, and outward of the Sun contains sections previously thought to belong to Cygnus and Perseus.

Almost half the stars in the Solar neighourhood, including the Sun, have orbits aligned with the Orion arm. Of the remainder, over a quarter (the Hyades stream) have orbits aligned with the Centaurus arm, and are currently crossing the Orion arm on their way outwards to rejoining the Centaurus arm later in their orbits. Another third are split between three streams, the Hercules, Sirius and Alpha Ceti streams, and have more eccentric orbits straddling both arms. These streams have been found to consist of older stars. Because spiral potential draws stars into approximately elliptical orbits with a range of eccentricities determined by pitch angle, it is expected that there will be a dearth of near circular orbits. This will lead to a well in the velocity distribution, giving an estimate the LSR.

\section{Metallicity and kinematics}\label{sec:metallicity}
\begin{figure}
	\centering
	\includegraphics[width=1\textwidth]{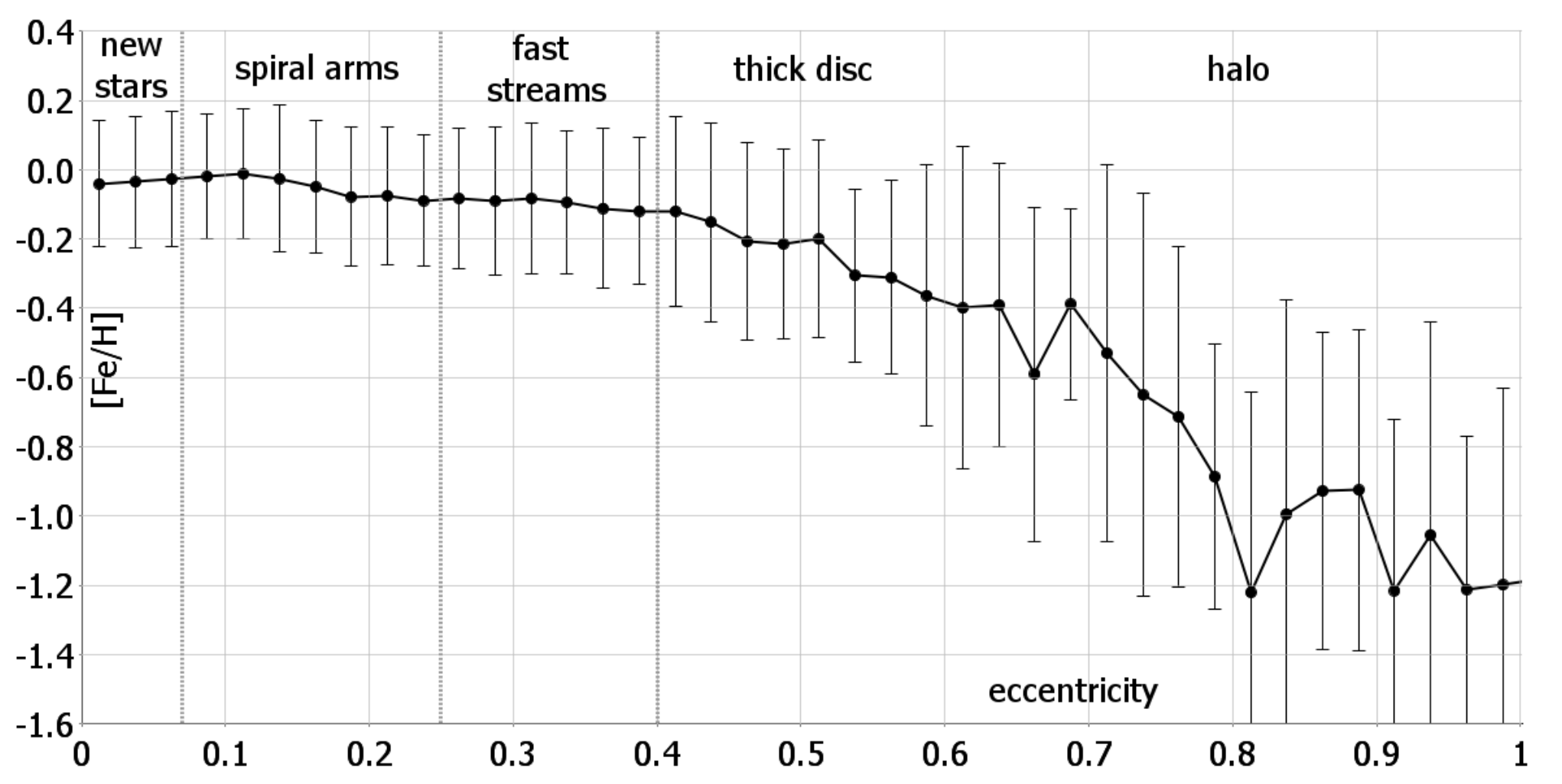}
	\caption{Mean metallicities binned by eccentricity, for 13\,954 Hipparcos stars in 40 bins. The error bars show the dispersion in each bin. Vertical grey dotted lines show the bounds of kinematic groups found in FA12.}
	\label{fig:ecc-fe}
	\end{figure}

Sch\"onrich, Binney \& Dehnen say in their abstract that the determination of the component of the LSR ``in the direction of Galactic rotation via Str\"omberg's relation is undermined by the metallicity gradient in the disc, which introduces a correlation between the colour of a group of stars and the radial gradients of its properties''. However, for 17\,227 stars with observed metallicities, the correlation coefficient between metallicity and Galactic radius in the solar neighbourhood is $ 0.020 $, and the metallicity gradient, $ 0.085 $ dex per kpc, is far smaller than random fluctuations (and actually of opposite sign to that expected). Using a smaller population and less accurate metallicities, Holmberg, Nordstr\"om and Andersen (2007) found $ -0.09 $ dex per kpc, which is of the expected sign but is still much less than the dispersion, $ 0.2 $ dex, in the Solar neighbourhood, and far too small to influence Str\"omberg's relation. Similarly, there is no indication in the data that stars at pericentre are less metal rich than stars at apocentre.

Nor is there any reason that metallicity should affect Str\"omberg's relation, since the supposed correlation is not with colour but with velocity dispersion. In a well-mixed distribution a correlation would appear whenever populations of stars with different eccentricities are selected, irrespective of metallicity. Each colour bin contains stars with a range of ages and kinematic properties. Metallicity has an influence on the membership of individual bins, but it does not alter the principle that redder bins tend to contain stars with more eccentric orbits.

There is a significant correlation, coefficient$ -0.43 $, between metallicity and the magnitude of velocity relative to the LSR. Much of the correlation is due to the thick disc and halo; after eliminating stars with eccentricities greater than $ 0.4 $ the correlation is $ -0.06 $. The correlation is shown in figure \ref{fig:ecc-fe} using 40 eccentricity bins for 13\,954 Hipparcos stars for which we have metallicities and complete kinematic information and parallaxes with better than 20\% errors. The error bars show the dispersion within each bin. The vertical grey dotted lines show the bounds of kinematic groups found in FA12.

The correlation results because older stars formed a time when the Galaxy was less metal rich. FA09b found that the Sirius, Alpha Ceti and Hercules streams consist in the main of stars older than 9 Gyrs, and identified that this is the time at which the Milky Way evolved into a bisymmetric spiral. Because they are older, fast moving streams are less metal rich, just as the thick disc is less metal rich than the thin disc. It follows that, in general, higher velocity stars tend to be less metal rich. Membership of fast streams is responsible for Parenago's discontinuity (1950, see FA09a, section 4), and is also responsible for the correlation found in Str\"omberg's relation. Figure 1 shows that metallicity in fast streams is slightly less than in the rest of the thin disc. As a result, stars of the same mass and age are slightly bluer, but this effect is small compared to the fact that these streams are substantially redder than the rest of the thin disc because they are older. 

Because of the long lives of red dwarfs, binning the population by colour does not produce populations of uniform age or of a single kinematic group. Redder bins contain a greater admixture of stars from fast streams and from the thick disc. There is no correlation for bins redder than $ B-V>4.3 $ mag (i.e. later than $ \sim $F3-4). The correlation in Str\"omberg's relation results from differences in velocity between spiral arms, fast moving streams, and the thick disc. It is not a progressive change and has little bearing on circular motion. In fact the non-working of Str\"omberg's relation was already seen in Dehnen \& Binney (1998). Firstly giants were rejected because they were held to follow different kinematics. Str\"omberg did use giants; the argument for the asymmetric drift relation depends only the use of populations with different eccentricities in a well-mixed distribution. If it works at all it should also work for giants. Secondly it was found that bins for early types do not lie on the line of regression. This data was also rejected, although, being recently formed from gas clouds, early types might be expected to show more nearly circular motion than the rest of the population (as is the case, Francis 2013).

The correlation between metallicity and velocity magnitude is a part of \linebreak Galactic spiral structure, and is not related to a metallicity gradient in the disc. It is well established that the inner galaxy is more metal rich. This is as one would expect, because there is more matter, more activity, larger stars and faster processes in the inner Galaxy. Metallicity depends on the number of generations of stars and supernovae the gas in a particular star forming region has been through. This is expected to vary in the disc in discrete amounts in an essentially random way. Hence, a smooth metallicity gradient is not expected, and, since orbits are not circular, a mix of metallicities should be found at any radius. There is thus no significance in the absence of a metallicity gradient at the solar position.

\section{Conclusions}\label{sec:Conclusions}
The velocity distribution of local stars is highly structured, and heavily biased towards membership of six major streams. This invalidates the traditional methods of calculating the LSR, based on Str\"omberg's asymmetric drift relation, for which a well-mixed distribution is required. We have found alternative indicators by examining the properties of the velocity distributions. We believe the best indicators are based on an observed minimum in the distribution which we believe represents circular motion. Although a minimum at the LSR can be seen as a consequence of heating of the disc, it is more properly understood from the relationship between streams and spiral structure (FA09b and FA12). We found agreement between the value of $U_0$ at the minimum and the mean, $\overline{U}$, for stars with $| W + 6.9 | \ge 40$ km\ s$^{-1}$, whose orbits are far enough out of alignment with the disc that they do not participate in bulk streaming motions.

Contrary to Sch\"{o}nrich, Binney \& Dehnen, there is no significant local correlation between metallicity and Galactic radius. What little slope there is locally is of opposite sign to the expected gradient to the Galactic centre. The absence of a local correlation is illustrative of mixing due the fact that stars do not follow circular, or near circular orbits. A correlation is found between metallicity and eccentricity, and is related to the age of fast moving streams.

It has been seen here that the improved database contains a deeper and more clearly defined well than was found in the previous study, and makes possible a better estimate of the LSR. We give as our best estimate of solar motion with respect to the LSR, $(U_0, V_0, W_0) = (14.1 \pm 1.1, 14.6 \pm 0.4, 6.9 \pm 0.1)$ km\ s$^{-1}$, in close agreement with the estimate found LSR from RAVE, $(U_0, V_0, W_0) = (14.9 \pm 1.7, 15.3 \pm 0.4, 6.9 \pm 0.1)$ km\ s$^{-1}$ which uses an almost entirely distinct population of stars (Francis, 2013). $W_0$ is found by restricting the population using Gaussian fitting. $V_0$ is found from the low frequency of stars in orbits with eccentricity less than 0.01, supported by the observed trough in the distribution for stars close to orbital extrema. $U_0$ is found by combining the estimate from the low frequency of stars in circular motion with the estimate from stars with orbits with a substantial inclination to the Galactic plane.

\section*{Data}
XHIP can be retrieved from the Centre de Donn\'{e}es astronomiques de Strasbourg (CDS Catalog V/137D).

\section*{Appendices}
\appendix

\section{Estimate of dispersion and eccentricity} \label{ApB}
Kepler's second law is an expression of conservation of angular momentum, and holds for planar orbits in any axisymmetric potential. Assuming a well-mixed distribution in which orbital eccentricities are low, the greater density of stars towards the centre of the Galaxy will not greatly affect calculation. If the number of stars on the outer part of the orbit (lagging the LSR) outnumbers the number on the inner part (leading the LSR) by 70:30, it is because the time on the outer part of the orbit is greater by roughly this ratio, so the velocity is less and the radial distance is greater by the square root of this ratio, i.e. by a factor greater than 1.5 (contradicting the assumption that orbital eccentricities are low). Then, for Galactic rotation $\theta_0 \approx 200 $ km\ s$^{-1}$, azimuthal velocities will range from $\sim 160 $ to $\sim 240 $ km\ s$^{-1}$ and dispersion in $ V $ will be about 40 km\ s$^{-1}$, around twice the observed value of 21 km\ s$^{-1}$ for disc stars.

If eccentricity, $ e $, is defined such that the ratio of distance to apocentre to distance to pericentre is given, as for a Keplerian orbit, by
\begin{equation}
\frac{1+e}{1-e} = \frac{R_{\mathrm{max}}}{R_{\mathrm{min}}}>\sim 1.5
\end{equation}
from which
\begin{equation}
e> \sim 0.2
\end{equation}
But the modal value of eccentricity in the solar neighbourhood is $ \sim 0.11 $ (figure \ref{fig:7}). The modal value of eccentricity is insensitive to the figure used for the LSR.

\section{Significance of the well} \label{ApA}
A simple estimate of the probability that the well at $ (U, V) = (-12.5, -14) $ km\ s$^{-1}$ seen in figure \ref{fig:UVdist} could arise by chance can be found by defining one hundred $2\times 2$ km$^2$\ s$^{-2}$ bins centred at even integer velocities in the ranges $ -24\le U \le -6$, $-24\le V \le -6 $ km\ s$^{-1}$. The bins contain 3 824 stars. The bin centred at (-12, -14) km\ s$^{-1}$ contains only 21 stars. For a uniform distribution (the null hypothesis) the number of stars in each bin is given by a binomial distribution $ \mathsf{B}(3 824, 0.01) $, for which the probability of finding 21 or fewer stars in a bin is 0.0017, giving a 99.8\% significance level. The use of a uniform probability distribution underestimates the statistical significance of the well, because there is generally a greater probability of finding data towards the centre of a distribution.

\end{document}